\documentclass[aps,pre,twocolumn,superscriptaddress,nofootinbib]{revtex4-2}

\usepackage{amssymb, amsmath, graphicx, xcolor}

\usepackage[colorlinks=true,linkcolor=blue,urlcolor=blue,citecolor=blue]{hyperref}

\usepackage[normalem]{ulem}

\begin{document}

\preprint{}

\title{Hybrid quantum-classical systems: statistics, entropy, microcanonical ensemble and its connection to the canonical ensemble}



\author{J. L. Alonso}
\email[]{alonso.buj@gmail.com}
\affiliation{Departamento de F\'isica Te\'orica, Universidad de Zaragoza, Zaragoza, Spain}
\affiliation{Centro de Astropartículas y Física de Altas Energías (CAPA), Zaragoza, Spain}
\author{C. Bouthelier-Madre}
\email[]{cbouthelier@ubu.es}
\altaffiliation[Present address: ]{Departamento de Matemáticas y Computación, Universidad de Burgos, Burgos, Spain}
\affiliation{Departamento de Matemáticas Fundamentales, Universidad Nacional de Educacion a Distancia, Madrid, Spain}
\author{A. Castro}
\email[]{acastro@uva.es}
\affiliation{Departamento de Física Teórica, Atómica y Óptica, Universidad de Valladolid, Valladolid, Spain}
\affiliation{Institute for Biocomputation and Physics of Complex Systems (BIFI), Zaragoza, Spain}
\author{J. Clemente-Gallardo}
\email[]{jcg@unizar.es}
\affiliation{Departamento de F\'isica Te\'orica, Universidad de Zaragoza, Zaragoza, Spain}
\affiliation{Centro de Astropartículas y Física de Altas Energías (CAPA), Zaragoza, Spain}
\author{J. A. Jover-Galtier}
\email[]{jorgejover@unizar.es}
\affiliation{Departamento de Matemática Aplicada, Universidad de Zaragoza, Zaragoza, Spain}
\affiliation{Instituto Universitario de Matemáticas y Aplicaciones (IUMA), Zaragoza, Spain}


\date{\today}

\begin{abstract}
We describe in detail a mathematical framework in which statistical ensembles of hybrid classical-quantum systems can be properly described. We show how a maximum entropy principle can be applied to derive the microcanonical ensemble of hybrid systems. We investigate its properties, and in particular how the microcanonical ensemble and its marginal classical and quantum ensembles can be defined for arbitrarily small range of energies for the whole system. We show how, in this situation, the ensembles are well defined for a continuum of energy values, unlike the purely quantum microcanonical ensemble, thus proving that hybrid systems translate properties of classical systems to the quantum realm. We also analyze the relation with the hybrid canonical ensemble by considering the microcanonical ensemble of a compound system composed of a hybrid subsystem weakly coupled to a reservoir and computing the marginal ensemble of the hybrid subsystem. Lastly, we apply the theory to the statistics of a toy model, which gives some insight on the different properties presented along the article.
\end{abstract}


\maketitle

\section{Introduction}

Equilibrium ensembles are fundamental in the description of statistical systems: the microcanonical ensemble describes isolated systems with fixed energy and number of particles, the canonical ensemble allows for interchange of energy with a thermal reservoir, and the grand canonical ensemble also permits variation in the number of particles of the system. These three equilibrium ensembles are mutually related, and their general characteristics are well known for both exclusively classical and exclusively quantum systems.

The situation is however less clear in the case of hybrid quantum-classical systems. In this article, we aim to investigate their statistical description and the properties of their equilibrium ensembles. Hybrid models incorporating both classical and quantum subsystems are useful in describing molecular and condensed matter systems \cite{Bornemann1996, Rapaport2004, Tully1998b}, which present two distinctive energy scales (nuclei and electrons). In the context of field theory, hybrid quantum-classical systems have also been considered as candidates to describe quantum matter fields interacting with a (classical) gravitational field, as a semiclassical approximation or even a fundamental theory \cite{Martin-Dussaud2019, Tilloy2019}. Applications of hybrid models include the description of the measurement process of quantum systems, by modeling the measurement device as a classical system coupled to the quantum system to be measured \cite{Buric2012, Buric2013, Diosi2014}, and the study of control systems \cite{Boghiu2024}.

An accurate and complete description of statistical system is fundamental in many applications, such as \emph{ab initio} modeling of molecules and materials and their numerical simulation methods at finite temperature \cite{Alavi1994, Alonso2021, Grumbach1994, Karasiev2014}. Some previous works have dealt with the description of hybrid statistical systems \cite{Alonso2020, Alonso2021, Goldstein2006, Hall2016, Mauri1993, Parandekar2005}, with a common approach of allowing a general and consistent description of interactions between quantum and classical ensembles. Choosing a consistent entropy function \cite{Alonso2020}, it is possible to provide a coherent mathematical framework for the description of statistical hybrid systems which determines the hybrid canonical ensemble as the equilibrium ensemble maximizing the entropy of the system under appropriate conditions \cite{Cohen-Tannoudji2020, Huang1963, Jaynes2003, Reichl1980, Vanslette2017}. This maximization property can also be used to identify the microcanonical ensemble, as we analyze in detail in the present paper. In principle, other equilibrium ensembles, such as the grand-canonical ensemble, can also be considered within this formalism. For instance, quite recently some first-principles proposals were introduced to consider the grand-canonical ensemble for quantum systems \cite{DelleSite2024, Reibl2025}. Using a similar approach, it should be possible to extend this construction to hybrid systems. That study is now in progress.

In \cite{Alonso2020} we have shown that, within a probability-based approach, the three cases, classical, quantum and hybrid, have, by the same right, their own entity (i.e. none of the three needs of the existence of the other two to be well-defined): their corresponding sample spaces are unequivocally characterized by the results of experiments. That is, we have imposed that Kolmogorov's axioms are satisfied \cite{Kolmogorov1956}. Other interesting articles partially work on the same line \cite{Barcelo2012, Camalet2025}.

In our description of hybrid statistics, exclusive events play a critical role in the appropriate definition of ensembles. Mutually exclusive events of a statistical system are those that can be differentiated by measuring observable quantities of the system \cite{Cafaro2013}. For a classical statistical system whose observables can be perfectly measured, mutually exclusive events are simple events that contain different points of the phase space. If however, observables are only measurable of a subset of phase space, or it is coarse grained (two different points too close together cannot be distinguished due to measurement resolution or because it is a non-physical quantity what makes them differ), then different situations arise with alternative definitions of mutually exclusive events.

In quantum mechanics, due to its probabilistic nature different approaches can be taken. On the one hand, statistical quantum systems can be treated in the same way as classical systems, identifying mutually exclusive events as simple events with different points of the phase space (in this case, a projective Hilbert space), with expectation values of observables as the tool to differentiate events. This approach was proposed in the mid of the 20th century \cite{Schrodinger1952, Khinchin1960} and developed later \cite{Bender2005, Brody2005, Naudts2006}, and it is used in fields such as quantum gravity \cite{Balasubramanian2024, Hawking1988, Kudler-Flam2021}. On the other hand, as we discussed in a previous article \cite{Alonso2015}, these models lack the extensivity properties that we expect from thermodynamic systems when interaction is either local or exhibits screening of large distance correlations. This limits significantly the use for systems where electromagnetic interaction plays an important role, such as molecular systems or materials.
 
Instead, other approaches take into account that measurements are probabilistic and only certain values can be obtained as results of measurements for each observable \cite{Hall2016, Neumann1955}. This implies that only those events that always provide different measurements for at least one observable can be considered to be mutually exclusive. Mutually exclusive events are thus linked to orthogonal eigenvectors of operators on the Hilbert space describing the states of the quantum system. In this article, we follow this approach and extend its properties to the description of statistical hybrid systems.

The article is organized as follows. Section \ref{sec:math} summarizes the main mathematical aspects used in our description of classical, quantum and hybrid statistical systems, introducing and defining the objects encoding the state of the different types of system. Then, Sec. \ref{sec:ensemble} describes the notion of hybrid entropy and how to use the maximum entropy formalism in the case of the microcanonical ensemble. In Sec. \ref{sec:lim0}, we analyze how the microcanonical ensemble is defined in the limit of small ranges of energies of the system, proving that the marginal classical and quantum ensembles share the classical property of being defined for almost any energy, unlike the typical quantum microcanonical ensemble. In Sec. \ref{sec:HybridCan} we explore the relation between the microcanonical and canonical hybrid ensembles, considering the microcanonical ensemble of a compound system composed of a hybrid subsystem weakly coupled to a reservoir. Section \ref{sexExample} discusses a simple example to obtain the microcanonical ensemble of a hybrid qubit while Sec. \ref{sec:conclusions} summarizes the main conclusions of the article and discusses potential future applications.

\section{Mathematical description of statistical systems}
\label{sec:math}

In probability theory, a statistical description of a system is based on the definition of a sample space as the set of possible configurations of the system. The subsets of the sample space are called events, and a probability measure is a real function on the set of events taking values between 0 and 1 and satisfying Kolmogorov axioms \cite{Kolmogorov1956}. This is achieved by considering a probability distribution, which involves a definition of a sample space and a probability measure defined on it. This is trivial to do in the classical world, but the quantum case is more subtle. In the following, we review the application of these concepts to our analysis of classical, quantum and hybrid systems.

\subsection{Classical states and statistics}

In classical statistical systems, the sample space is identified with the phase space $M_C$ of the system. States $\xi \in M_C$ represent the values of all the variables determining the configuration of the system (e.g., positions and velocities in a mechanical system). In classical statistics, simple events are mutually exclusive: if two simple events $A_1 = \{\xi_1\}$, $A_2 = \{\xi_2\}$ contain different points, $\xi_1 \neq \xi_2$, then conditional probabilities of $A_1$ with respect to $A_2$ and viceversa are zero, i.e., the probability of detecting the system in state $\xi_1$ after it has been prepared in $\xi_2$ is zero, and viceversa.

An important element in the formalism is the description of observables. Any physical quantity that can be measured on the system (energy, temperature, position of a particle, etc.) is considered an observable, and thus is represented by a function on $M_C$. Mathematically, the set of classical observables $\mathcal{O}_C$ is a Poisson algebra\footnote{This requires the existence of a Poisson bracket on the algebra of functions, which allows us to define Hamiltonian dynamics. The Poisson bracket is just a skew-symmetric bidifferential operator that satisfies the Jacobi identity when acting on three arbitrary functions. See Ref. \cite{Abraham1978} for details.} of integrable smooth real valued functions on $M_C$, with an inner product defined by integration
\begin{equation}
\label{eq:inProdOC}
(a,b) = \int_{M_C} a(\xi) b(\xi) d \mu_C, \quad a, b \in \mathcal{O}_C,
\end{equation}
with $d \mu_C$ the volume element on $M_C$. Such a set can be considered (see \cite{BouthelierMadre2023}) to be the subset of real elements of the $C^*$ algebra of complex functions with compact domain, and the supremum norm:
\begin{equation}
\mathcal{O}_C=\left\{ f:M_C\to \mathbb{C} \; \left | \, \| f\|= \mathrm{sup} |f| < \infty \right . \right  \}.
\end{equation}

Statistical ensembles represent the uncertainty in the knowledge of the actual state of the system, which translate into uncertainty on the measured values of observables. For this reason, statistical ensembles are represented by distributions (also called generalized functions): linear continuous functions on the algebra of observables, i.e. elements in the dual space $\mathcal{O}^*_C$ with respect to the inner product \eqref{eq:inProdOC}. The set of probability distributions $\mathcal{P}_C \subset \mathcal{O}^*_C$ is determined by the necessary conditions to provide a physical meaning of probability to a distribution $F \in \mathcal{P}_C$ written as a measure in terms of a probability density with respect to the volume element $d\mu_C$:
\begin{equation}
\label{eq:normFC}
\int_U F(\xi) d\mu_C  \geq 0, \ \forall \mbox{ open } U \subset M_C, \ \int_{M_C} F(\xi) d\mu_C = 1.
\end{equation}
This definition of statistical ensembles in term of distributions allows for nonsmooth functions and Dirac $\delta$ distributions to also be considered.

For an ensemble described by a probability distribution $F \in \mathcal{P}_C$, the average value of any observable $a \in \mathcal{O}_C $ is defined naturally as the action of $F$ on $a$:
\begin{equation}
\langle a \rangle = F[a] = \int_{M_C} F(\xi) a(\xi) d\mu_C .
\end{equation}

For any subset $U \subset M_C$, the probability of finding the system at a state in $U$ is
\begin{equation}
p_C(U; F) = \int_U F(\xi) d\mu_C.
\end{equation}
A pure state in which the system is known to be in a certain state $\xi_0 \in M_C$ is represented by the Dirac $\delta$ distribution $\delta \left( \xi- \xi_0 \right) \in \mathcal{P}_C$.

\subsection{Quantum states and statistics}

The objects presented in the description of classical statistical system are general, and an analogous approach has to be developed to describe quantum statistics. However, the first difference appears in the description of sample spaces and mutual exclusive events, which is nontrivial compared to the classical case due to the linear and probabilistic nature of quantum mechanics.

In the Schr\"odinger picture of quantum mechanics, the pure states of a quantum system are described by vectors in a complex Hilbert space $\mathcal{H}$. The relation between states and vectors $|\psi\rangle \in \mathcal{H}$, however, is not unique, and any two proportional nonzero vectors describe the same state. As a consequence, the space of pure states $M_Q$ of a quantum system can be characterized as the projective Hilbert space, which is itself diffeomorphic to the space $\mathcal{P}_Q^1$ of rank-1 projectors onto its Hilbert space $\mathcal{H}$. Using Dirac notation, the space $M_Q$ is described as
\begin{equation}
M_Q = \mathcal{P}_Q^1 = \left\{ \pi = \frac{|\psi\rangle \langle \psi|}{\langle \psi|\psi \rangle} \mid |\psi\rangle \in \mathcal{H} - \{0\} \right\}.
\end{equation}
We will use $\pi$ to denote rank-1 projectors, i.e. pure states, and reserve $\rho$ for mixed states, which clarifies some formulas below.

Observables of a quantum system are represented by self-adjoint operators on $\mathcal{H}$, which constitute the real (with respect to the canonical involution) part  of a unital $C^*$ algebra $\mathcal{O}_Q$ \cite{BouthelierMadre2023, Carinena2015book, Strocchi2005}. In the case of finite-dimensional quantum systems, this set corresponds simply to the algebra of Hermitian matrices of appropriate dimension. Unlike classical systems, the result of a measurement on a quantum system is not deterministic, and in general there exists uncertainty in the results of the measurement process even if the state of the system is fully determined. This is characterized by the following property: if a system is in a state $|\psi_1 \rangle \in \mathcal{H} - \{0\}$, then the probability $p_{12}$ of finding the system in the state $|\psi_2 \rangle \in \mathcal{H} - \{0\}$ is \cite{Penrose1979}
\begin{equation}
    \label{eq:p12}
p_{12} = \frac{\left| \langle \psi_1 | \psi_2 \rangle \right|}{\| \psi_1 \|^2 \|\psi_2\|^2} = {\rm Tr} (\pi_1 \pi_2), \quad \pi_i = \frac{|\psi_i\rangle \langle \psi_i|}{\langle \psi_i|\psi_i \rangle}, \ i = 1,2.
\end{equation}
In words, the probability of measuring a system to be detected in a state when we have prepared it, with absolute certainty, in a different state may be different from zero.

Due to this intrinsically probabilistic nature of quantum mechanics, a definition of a sample space as the whole set of distinct states $M_Q$ (as it was done in the classical case), is problematic. 
Given any probability measure $p$ defined on $M_Q$, if we consider two different simple events $\lbrace \pi_1\rbrace$ and $\lbrace \pi_2\rbrace$, they are mutually exclusive, since 
$\lbrace \pi_1\rbrace \cap \lbrace \pi_2\rbrace = \emptyset$ and therefore $p(\lbrace \pi_1\rbrace \cap \lbrace \pi_2\rbrace) = 0$.
This seems to contradict the previous statement about nonzero probabilities for different states, but it does not: a probability measure defined on $M_Q$ must be interpreted as a
{\em preparation} measure. Hence, $p(\pi_1)$ is the probability that the system has been prepared in state $\pi_1$. If $p(\pi_1)=1$, then there is obviously zero probability that it has been prepared in a different state $\pi_2$.
But then, as shown in Eq. \eqref{eq:p12}, there is a nonzero probability of being
{\em detected} or {\em measured} in that different state $\pi_2$, if it is not orthogonal to $\pi_1$.

In this work, our approach will be the following \cite{Hall2016, Neumann1955}: we will say that two states $\pi_1$, $\pi_2$ are mutually exclusive states (MESs) if ${\rm Tr}(\pi_1 \pi_2) = 0$. Thus, we are considering a {\em measuring probability}. To properly define it, one must consider, as possible sample spaces, subsets of states in $M_Q$ whose associated simple events are mutually exclusive in that definition.
 
How to represent a statistical state of the quantum system, then?  The state of this statistical system can be always represented by a density matrix of the given system. This is in fact  the consequence of  Gleason's theorem \cite{Gleason1975}: any measure of a quantum system (with Hilbert space's dimension larger than 2) can be written as expectation values with respect  to a unique  self-adjoint, positive and trace-class operator, with trace equal to one. This defines a density matrix of said system, which in this formalism can be identified with a linear functional on $\mathcal{O}_Q$ and written as a convex combination of rank-1 projectors \cite{Strocchi2005}. The corresponding set of density matrices $\mathcal{P}_Q$ contains those linear operators $\rho$ on $\mathcal{H}$ such that
\begin{equation}
\rho^\dagger = \rho, \quad
\rho \geq 0, \quad
\mathrm{Tr}(\rho) = 1.
\end{equation}
Notice that $\mathcal{P}_Q^1 \subset \mathcal{P}_Q$, and that the convex combination of any two density matrices is also a density matrix. Expectation values for observables are
\begin{equation}
\langle A \rangle = \mathrm{Tr}( A \rho), \quad  \rho \in \mathcal{P}_Q, \ A \in \mathcal{O}_Q.
\end{equation}
Thus we see that the state of the system determines a linear functional on the $C^*$ algebra of observables, i.e. an element of the dual space $\mathcal{O}_Q^*$.

In principle it is also possible to define mixed quantum states as probability distributions on the set of pure states $M_Q$, i.e., as a positive-definite generalized function ${F_Q:M_Q\to \mathbb{R}}$ normalized to one. This implies considering $M_Q$ as a sample space, analogously to the classical case. In this situation, the expectation value of an observable $\hat A\in \mathcal{O}_Q$ can be written as
\begin{equation}
\langle \hat A \rangle=\int_{M_Q} d\mu_Q F_Q(\pi) f_A(\pi),
\end{equation}
where $f_A(\pi)=\mathrm{Tr}(\hat A \pi)$ and with $d \mu_Q$ the volume element on $M_Q$. Nonetheless, Gleason's theorem ensures that only the first moment of the distribution is relevant and 
\begin{equation}
\langle \hat A \rangle=\mathrm{Tr} (\rho \hat A),
\end{equation}
where the density matrix $\rho$ which corresponds to the first moment of $F_Q$ can be written as
\begin{equation}
\rho= \int_{M_Q} d\mu_Q F_Q(\pi) \pi.
\end{equation}
In this way, independence of the different events is captured in the eigenspaces of the density matrix. Nonetheless, notice that two different probability distributions $F_Q$ are physically equivalent as long as they have the same first moment. No measurement can distinguish between the two distributions and hence they must correspond to the same physical state.

Quantum dynamics is linear on the space of states and therefore can be written as a master equation for the density matrix (i.e., the first moment of the distribution). We will see below that the situation is radically different for hybrid systems.

The relation between pure and mixed states is crucial for the following sections. The spectral decomposition theorem shows that any density matrix can be described as a convex combination of projectors of an orthonormal basis on $\mathcal{H}$; in other words, any ensemble can be written in terms of elements of a maximal set of MESs. Notice that any other state not in the set is not immediately excluded from the ensemble (after all, is not mutually exclusive with those in the set). Thus, the probability of finding the quantum system at any state $\pi \in M_Q$ (not only those MESs considered above) for an ensemble described by a density matrix $\rho \in \mathcal{P}_Q$ is
\begin{equation}
p_Q(\pi; \rho) = \mathrm{Tr}(\rho \pi).
\end{equation}

Notice also that, when we choose to represent the state as a probability density over $M_Q$, all distributions having the same first moment, from Gleason theorem,  are represented by the same density matrix. Therefore, they associate the same expectation value to any physical observable and they are hence physically indistinguishable. For example, an equiprobable distribution over the whole state space $M_Q$ of a two-level quantum system is indistinguishable from a distribution of any pair of MESs with equal probability, all of which define the density matrix $\frac{1}{2} \mathbb{I}$.

\subsection{Hybrid classical-quantum states and statistics}

Hybrid quantum-classical systems are composed of two coupled subsystems, one of them classical and the other one quantum \cite{Alonso2020}. Its phase space is therefore $M_H = M_C \times M_Q$, whose elements are pairs $(\xi, \pi)$ of classical and quantum states, as defined in previous sections.

In the following, it will be useful to describe $M_H$ as a trivial fiber bundle, with $M_C$ as the base space and $M_Q$ acting as fibers. This facilitates the incorporation of the particular characteristics of quantum systems into the description of hybrid statistics.

\begin{figure}[h]
\centering
\includegraphics[width=0.45\textwidth]{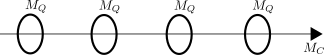}
\caption{Hybrid manifold $M_H$ as a fiber bundle.}
\label{fig:FB}
\end{figure}

Associated with that bundle, we can consider the bundle over $M_C$ having as fiber the $C^*$ algebra of quantum operators $\mathcal{O}_Q$. With this description, observables $\hat{\mathcal{O}}_H$ can be considered to correspond to the set of sections of that bundle, i.e. the set of objects of the form $\hat{A}(\xi)$, where for each $\xi\in M_C$, $\hat{A}(\xi)$ defines an element of $\mathcal{O}_Q$, i.e., a well-defined quantum observable on $M_Q$. The set of hybrid operators $\mathcal{O}_H$ becomes isomorphic to the $C^*$ algebra $\mathcal{O}_C\otimes \mathcal{O}_Q$ (see Ref. \cite{BouthelierMadre2023} for the technical details concerning the definition of a suitable norm on the space of hybrid operators).

Statistical hybrid systems can be defined, by mimicking the classical and quantum cases, as probability distributions on $M_H$, i.e., defined by a positive-definite function $F_{QC}:M_H\to \mathbb{R}$, normalized to one. Expectation values are obtained as
\begin{equation}
\langle \hat A (\xi) \rangle= \int_{M_H}d\mu_H F_{QC}(\xi,\pi)f_{A}(\xi,\pi),
\end{equation}
where now
\begin{equation}
f_A(\xi,\pi)=\mathrm{Tr} (\hat A (\xi)\pi), \quad d\mu_H = d \mu_C d \mu_Q.
\end{equation}
As in the other cases, statistical ensembles should be described as elements in $\hat{\mathcal{O}}_H^*$. Let us provide, briefly, an appropriate context for such objects. The interested reader can find more details in Refs. \cite{Alonso2020,BouthelierMadre2023}. Description of MESs in hybrid systems can be determined following the description of MESs in purely classical and quantum systems. Two points $ (\xi_1, \pi_1), (\xi_2, \pi_2) \in M_H$ define MESs if and only if $\xi_1 \neq \xi_2$ or $\pi_1 \pi_2 = \pi_2 \pi_1 = 0$. Thus, hybrid systems share mutual-exclusivity properties with both classical and quantum systems. Let us assume that a probability distribution over $M_H$ is given by a probability density $F_{QC}$. If one is only interested in the classical variable and defines the corresponding marginal probability, then the ensemble should behave as a classical statistical system, i.e. as a distribution in $\mathcal{P}_C$. However, for a fixed value of $\xi \in M_C$, the conditional distribution should be just quantum, and hence from Gleason theorem described by a density matrix in ${P}_Q$. This justifies the following description of hybrid statistical ensembles, from now called hybrid density matrices:
\begin{equation}
\label{eq:defRhoH}
\hat{\rho} (\xi) = F_C (\xi) \tilde{\rho}_{Q,\xi}, \quad \forall \xi \in M_C.
\end{equation}
where $F_C \in \mathcal{P}_C$ is the classical marginal distribution and $\tilde{\rho}_{Q,\xi} \in \mathcal{P}_Q$ represents the quantum conditional probability for the quantum state if we assume that the classical subsystem is in the state $\xi$. For every $\xi \in M_C$, as long as $F_C (\xi)$ is finite and different from zero, $\hat{\rho}(\xi)$ is a positive-definite Hermitian operator on $\mathcal{H}$. Thus, hybrid density matrices $\hat{\rho}$ are elements in the set $\hat{\mathcal{P}}_H = \mathcal{P}_C \otimes \mathcal{P}_Q$, which is itself a subset of $\hat{\mathcal{O}}_H^*$, whose action on hybrid observables defines their expectation values:
\begin{equation}
\langle \hat{A} \rangle = \int_{M_C} \mathrm{Tr}(\hat{A}(\xi) \hat{\rho}(\xi)) d \mu_C, \quad
\hat{A} \in \hat{\mathcal{O}}_H.
\end{equation}

Marginal distributions of each subsystem can be computed by integrating or tracing out the relevant degrees of freedom:
\begin{equation}
F_C (\xi) = \mathrm{Tr}(\hat{\rho}(\xi)), \quad \rho_Q = \int_{M_C} \rho(\xi) d \mu_C.
\end{equation}
Notice that $\rho_Q \neq \tilde{\rho}_{Q,\xi}$, although both are purely quantum density matrices: $\rho_Q$ is the quantum marginal distribution of the original hybrid distribution, while $\tilde{\rho}_{Q,\xi}$ represents the conditional distribution for a given $\xi\in M_C$. However, $F_C$ is the same marginal distribution used in Eq. \eqref{eq:defRhoH}, which imposes the following conditions on the hybrid density matrix, derived from Eq. \eqref{eq:normFC}:
\begin{equation}
\label{eq:normRH}
\begin{gathered}
\int_U \mathrm{Tr}(\hat{\rho}(\xi)) d \mu_C \geq 0, \ \forall \mbox{ open } U \subset M_C; \\
\int_{M_C} \mathrm{Tr}(\hat{\rho}(\xi)) d \mu_C = 1.
\end{gathered}
\end{equation}

Notice that, as it was explained in Ref. \cite{Alonso2023b}, this hybrid density matrix represents the first quantum moment of the distribution $F_{QC}$ since
\begin{equation}
\rho(\xi)=\int_{M_Q}d\mu_Q F_{QC}(\xi,\pi)\pi.
\end{equation}
Therefore, this is  the generalization to the hybrid case of the definition of the density matrix seen for the quantum one. Hybrid density matrices have already been described in previous works, such as those by Aleksandrov \cite{Aleksandrov1981}, Kapral and Ciccotti \cite{Kapral1999} and our group \cite{Alonso2020, Alonso2021, BouthelierMadre2023, Alonso2023b} among many others.

\section{Description of the hybrid microcanonical ensemble}
\label{sec:ensemble}

Several relevant statistical ensembles exist in the literature. One of them is the canonical ensemble, which can be defined as the equilibrium ensemble for a closed system at constant temperature; i.e. with constant number of particles but with energy interchange between the system and a thermal reservoir. In a similar fashion, the microcanonical ensemble is defined as the equilibrium ensemble for an isolated system with constant energy \cite{Cohen-Tannoudji2020, Huang1963, Reichl1980}. Other ensembles, such as the grand canonical ensemble, could be defined with different conditions.

The equilibrium condition for these ensembles can be imposed in terms of the entropy. According to the principles of thermodynamics, an equilibrium distribution will maximize the entropy of the system. Thus, a maximum entropy computation can be performed to determine the equilibrium distributions, as we showed in Ref. \cite{Alonso2020} for the hybrid canonical ensemble.

In this section, we review entropy definitions presented in Ref. \cite{Alonso2020}, to apply the maximum entropy principle to determine the microcanonical ensemble of hybrid systems.

\subsection{Entropy definitions}

Consider a classical system with phase space $M_C$ whose states are governed by a certain probability distribution. The Gibbs entropy associated to such distribution is
\begin{equation}
\label{eq:SG}
S_G[F] = -\int_{M_C} F(\xi) \log(F(\xi)) d \mu_C, \quad F \in \mathcal{P}_C,
\end{equation}
taking the Boltzmann constant equal to 1. The classical microcanonical, canonical and grand canonical ensembles can be easily obtained by maximizing this entropy under the appropriate constraints \cite{Reichl1980}. It is relevant for our discussion the fact that classical entropy is an extensive property. Although the Gibbs entropy, also known as the differential entropy, lacks some desirable properties, such as invariance under change of variables and positivity, it is nonetheless appropriate for our discussion of hybrid systems.

In quantum mechanics, probability distributions are replaced by density matrices, whose entropy can be determined by von Neumann's formula \cite{Neumann1955, Penrose1979}:
\begin{equation}
\label{eq:SN}
S_N[\rho] = -\mathrm{Tr}( \rho \log (\rho)), \quad \rho \in \mathcal{P}_Q,
\end{equation}
with $\hbar = 1$. As before, maximization of this entropy under the appropriate constraints allows for the definition of equilibrium ensembles \cite{Reichl1980,Cohen-Tannoudji2020}.

In the case of hybrid systems, the entropy function with extensivity properties for these ensembles is
\cite{Alonso2020}
\begin{equation}
\label{eq:SH}
S_H[\hat{\rho}] = -\int_{M_C} \mathrm{Tr} \left( \hat{\rho}(\xi) \log \hat{\rho}(\xi) \right) d \mu_C ,\quad \hat{\rho} \in \mathcal{P}_H.
\end{equation}
Notice that, within the bundle picture of Fig. \ref{fig:FB}, the hybrid entropy computes for each $\xi \in M_C$ the entropy of the quantum subsystem by von Neumann's formula \eqref{eq:SN}; if the result is interpreted as the classical entropy for a system with classical state $\xi$, then the extensivity property motivates the integration over all possible values of $\xi$, resulting in \eqref{eq:SH}.

The hybrid density \eqref{eq:SH} reduces to Eq. \eqref{eq:SG} or Eq. \eqref{eq:SN} for those ensembles for which one of the subsystems is pure.

\subsection{Hybrid microcanonical ensemble from a maximum entropy principle}

Classical and quantum microcanonical ensembles are known to be obtained by maximizing the associated entropies over systems with constant energy $E$. The obtained ensembles are equiprobable distributions of states with energy $E$. In this section, we prove that an analogous result is obtained when the hybrid entropy \eqref{eq:SH} is maximized under such assumption.

To apply the maximum entropy principle, let us consider the hybrid Hamiltonian observable $\hat{H} \in \hat{\mathcal{O}}_H$, determining the energy of the system. As described above, the hybrid Hamiltonian defines, at each $\xi \in M_C$, a quantum observable $\hat{H}(\xi) \in \mathcal{O}_Q$ for which the eigenvalue equation can be stated. Assuming a discrete finite spectrum of the Hamiltonian at every $\xi$, the eigenvalue equation is
\begin{equation}
\begin{gathered}
\hat{H}(\xi) |u_{i,j}^\xi\rangle = e_i^\xi |u_{i,j}^\xi\rangle, \quad \langle u_{i,j}^\xi |u_{i,j}^\xi\rangle = 1, \\
i = 0,1,2,\ldots, n ; \ j = 1,2, \ldots, N_i^\xi; \ \xi \in M_C,
\end{gathered}
\end{equation}
where $N_i^\xi$ denotes the multiplicity of the $i$th eigenvalue at $\xi$. The $\xi$-dependence of both eigenvalues and eigenvectors is made explicit, as both can, independently, vary along with $\xi$. The resulting eigenvectors define the adiabatic basis $b_{\hat{H}}(\xi) = \{|u_{i,j}^\xi\rangle\}$, a basis of $\mathcal{H}$ that is different for every classical state $\xi$.

A maximal set of MESs on $M_H$ can be constructed from the adiabatic basis as follows
\begin{equation}
\label{eq:MEEs_H}
\mathcal{B}_{\hat{H}} = \left \{ (\xi, \pi_{i,j}^\xi) \in M_H \mid \pi_{i,j}^\xi = |u_{i,j}^\xi\rangle \langle u_{i,j}^\xi|, |u_{i,j}^\xi\rangle \in b_{\hat{H}}(\xi) \right \}.
\end{equation}
By Eq. \eqref{eq:defRhoH}, hybrid densities can be fully described by determining the contribution of each element in $\mathcal{B}_{\hat{H}}$. To describe the hybrid microcanonical ensemble, only those elements in $\mathcal{B}_{\hat{H}}$ with energy in a certain interval $[E, E + \epsilon]$ should be allowed, which results in the following subset:
\begin{equation}
\label{eq:subMESs}
\mathcal{A}_{\hat{H}}^{E, \epsilon} = \left\{ \left( \xi, \pi_{i,j}^\xi \right) \in \mathcal{B}_{\hat{H}} \mid \mathrm{Tr} \left( \hat{H}(\xi) \pi_{i,j}^\xi \right) \in [E, E + \epsilon] \right\}.
\end{equation}
Let $\iota(\xi)$ denote the (possibly empty) set of pairs $(i,j)$ such that $(\xi, \pi_{i,j}^\xi) \in \mathcal{A}_{\hat{H}}^{E, \epsilon}$. Physically, this set indicates which elements of the adiabatic basis at $\xi$ have energy within the allowed interval.  Restricting Eq. \eqref{eq:defRhoH} to Eq. \eqref{eq:subMESs} gives the following expression for the hybrid density matrix describing the hybrid microcanonical ensemble at every $\xi \in M_C$:
\begin{equation}
\label{eq:propRHmC}
\hat{\rho}_\mu (\xi) = \left\{ \begin{aligned}
& \sum_{(i,j) \in \iota(\xi)} p_{i,j}^\xi \pi_{i,j}^\xi, && \iota(\xi) \neq \emptyset, \\
& 0, && \iota(\xi) = \emptyset.
\end{aligned} \right.
\end{equation}
The coefficients $p_{i,j}^\xi$ represent the probabilities associated to states $(\xi,\pi_{i,j}^\xi)$, and have to be determined by a maximum entropy principle. To be globally defined, we will consider that these coefficients exist for every $\xi \in M_C$, being equal to zero if the the corresponding hybrid state does not have energy within the allowed interval, i.e. $p_{i,j}^\xi = 0$ if either $\iota(\xi) = \emptyset$ or $(i,j) \notin \iota(\xi)$. With these values, the hybrid density matrix \eqref{eq:propRHmC} can be simply written as
\begin{equation}
\label{eq:propRHmC2}
\hat{\rho}_\mu (\xi) = \sum_{i=0}^n \sum_{j=1}^{N_i^\xi} p_{i,j}^\xi \pi_{i,j}^\xi.
\end{equation}
For the sake of simplicity, we will assume in the following that these coefficients can be defined by a generalized function on $M_H$. In such a case, the classical marginal distribution can be obtained by taking the trace of Eq. \eqref{eq:propRHmC2} to define
\begin{equation}
F_C(\xi) = \sum_{i=0}^n \sum_{j=1}^{N_i^\xi} p_{i,j}^\xi,
\end{equation}
with the normalization condition \eqref{eq:normRH}:
\begin{equation}
\int_{M_C} \sum_{i=0}^n \sum_{j=1}^{N_i^\xi} p_{i,j}^\xi d \mu_C = 1 .
\end{equation}

Considering the hybrid entropy \eqref{eq:SH} and imposing the normalization condition \eqref{eq:normRH} as a constraint with a Lagrange multiplier $\lambda$ gives as a result the following functional:
\begin{equation}
\begin{aligned}
\widetilde{S}_H[\hat{\rho}] = & S_H [\hat{\rho}] + \lambda \left( \int_{M_C} \mathrm{Tr} ( \hat{\rho} (\xi)) d \mu_C -1 \right ) \\
= & \int_{M_C}\mathrm{Tr} \left(  \lambda \hat{\rho} (\xi) - \hat{\rho}(\xi) \log \hat{\rho}(\xi) \right) d \mu_C -\lambda.
\end{aligned}
\end{equation}
Evaluating the functional $\widetilde{S}_H$ on the proposal for the microcanonical ensemble\eqref{eq:propRHmC}, and as $\mathrm{Tr}(\pi) = 1$ for every rank-1 projector $\pi \in M_Q$, the expression of $\widetilde{S}_H$ depends only on the probabilities $P_{(\xi,\pi)}$, with integration only over $\xi$ values with contribution to $\hat{\rho}_\mu$:
\begin{equation}
\begin{gathered}
\widetilde{S}_H[\hat{\rho}_\mu] =
\int_{M'_C} \sum_{i=0}^n \sum_{j=1}^{N_i^\xi} \left( \lambda p_{i,j}^\xi - p_{i,j}^\xi \log p_{i,j}^\xi \right)  d \mu_C, \\
M'_C = \{ \xi \in M_C \mid \mathfrak{a}(\xi) \neq \emptyset \}.
\end{gathered}
\end{equation}
A maximum of $\widetilde{S}_H$ is achieved by requiring its variation $\delta \widetilde{S}_H$ to be zero, which depends on the variations $\delta p_{i,j}^\xi$ of the probabilities $p_{i,j}^\xi$, defined for all $\xi \in M_C$ as
\begin{equation}
\delta \widetilde{S}_H[\hat{\rho}_\mu] = 0 \Rightarrow
\sum_{i=0}^n \sum_{j=1}^{N_i^\xi} \left( \lambda - 1 - \log p_{i,j}^\xi \right) \delta p_{i,j}^\xi = 0.
\end{equation}
The variations of those probabilities fixed to zero are assumed to always be zero; for the remaining probabilities, i.e. for all $(i,j) \in \iota(\xi) \neq \emptyset$, each of the coefficients of the variations have to be zero:
\begin{equation}
\lambda - 1 - \log p_{i,j}^\xi = 0 \Rightarrow p_{i,j}^\xi = \exp \left(\lambda -1 \right), \quad \forall (i,j) \in \iota(\xi).
\end{equation}
That is, all possible states $(\xi,\pi_{i,j}^\xi) \in \mathcal{A}_{\hat{H}}^{E, \epsilon}$ have the same probability. We thus conclude that the equilibrium hybrid density matrix for the microcanonical ensemble is an equiprobable distribution of all hybrid MESs with feasible energy in the interval $[E, E + \epsilon]$:
\begin{equation}
\label{eq:HmC}
\hat{\rho}_\mu (\xi) = \frac{1}{\Omega} \left\{ \begin{aligned}
& \sum_{(i,j) \in \iota(\xi)}\pi_{i,j}^\xi, && \iota(\xi) \neq \emptyset, \\
& 0, &&\iota(\xi) = \emptyset.
\end{aligned} \right.
\end{equation}
with $\Omega$ a normalization factor such that $\int_{M_C} \mathrm{Tr} ( \hat{\rho}_\mu (\xi)) d \mu_C =1$.
This factor can be computed as:
\begin{equation}
\label{eq:Omega}
\Omega = \int_{M_C}d\mu_C \vert \iota(\xi)\vert\,,
\end{equation}
where $\vert\iota(\xi)\vert$ is the cardinal or number of elements in the set $\iota(\xi)$. Notice that, disregarding the constant required to fix the dimensionality, $\Omega$ corresponds to the microcanonical partition function and allows us to obtain magnitudes such as the entropy as
\begin{equation}
S_H(\hat{\rho}_\mu) = \log\left(\Omega \right)\,.
\end{equation} 

The microcanonical ensemble \eqref{eq:HmC} is not only coherent with the mathematical formalism presented, but reproduces the expected characteristics of the microcanonical ensemble. As in the purely classical and quantum cases, the hybrid microcanonical ensemble can also be interpreted as an equiprobable distribution of MESs with energy between $E$ and $E+\epsilon$. Additionally, when the hybrid microcanonical ensemble is considered in the purely quantum or classical limits, it correctly reproduces the corresponding ensembles:
\begin{itemize}
\item The purely classical limit of a hybrid system can be achieved by considering that its quantum subsystem has a single energy level; thus, $\mathcal{H} = \mathbb{C}$ and therefore $M_Q = \{1\}$ and $\mathcal{O}(\mathcal{H}) = \mathbb{R}$. Hybrid observables are thus scalar functions on $M_C$, as they associate a quantum observable (i.e. a real number) to every $\xi \in M_C$. Identifying the hybrid Hamiltonian as a scalar function $\hat{H}(\xi) = h(\xi)$, we simply have
\begin{equation}
\begin{gathered}
\mathcal{B}_{\hat{H}} = \{(\xi,1) \in M_H \} = M_H, \\
\mathcal{A}_{\hat{H}}^{E,\epsilon} = \{(\xi,1) \in M_H \mid h(\xi) \in [E, E + \epsilon] \},
\end{gathered}
\end{equation}
and the microcanonical ensemble \eqref{eq:HmC} is
\begin{equation}
\label{eq:class_mC}
\hat{\rho}_\mu (\xi) = \left\{ \begin{aligned}
& \frac{1}{\Omega} , && h(\xi) \in [E, E + \epsilon], \\
& 0, && \mbox{otherwise},
\end{aligned} \right.
\end{equation}
which coincides with the classical microcanonical ensemble, in which any classical state with energy within the allowed interval $[E, E + \epsilon]$ is equally probable. 

\item The purely quantum limit of a hybrid system is obtained when the classical system has a single possible state, $M_C = \{\xi_0\}$. The hybrid Hamiltonian is effectively independent of the classical variables and is simple a quantum observable, $\hat{H} (\xi_0) = H \in \mathcal{O}(\mathcal{H})$, whose eigenstates form an orthonormal basis $b_H = b_{\hat{H}} (\xi_0)$ of $\mathcal{H}$; hence
\begin{equation}
\mathcal{B}_{\hat{H}} = \{ \xi_0 \} \times \{ \pi_{i,j} = |u_{i,j}\rangle \langle u_{i,j}| \mid |u_{i,j}\rangle \in b_H\}.
\end{equation}
The microcanonical ensemble is thus determined by these eigenstates. Assume the spectrum ${\rm spec}(H)$ of the Hamiltonian is discrete and the allowed interval $[E, E + \epsilon]$ contains a single eigenvalue $e_i$ with multiplcity $N_i$ and a basis $|u_{i,j}\rangle$ of its fundamental subspace; then, the corresponding microcanonical ensemble \eqref{eq:HmC} becomes
\begin{equation}
\label{eq:quant_mC}
\hat{\rho}_\mu (\xi_0) = \frac{1}{N_i} \sum_{j} \pi_{i,j},
\end{equation}
The quantum limit of the hybrid microcanonical ensemble coincides with the usual microcanonical ensemble for quantum systems\cite{Cohen-Tannoudji2020, Reichl1980}, where the density matrix is a convex combination of eigenstates of a certain energy with same weights.
\end{itemize}

Notice that the hybrid microcanonical ensemble \eqref{eq:HmC} preserves the linearity properties of its quantum subsystem. That is, the probability of any state not mutually exclusive with those in $\mathcal{A}_{\hat{H}}^{E, \epsilon}$ is in general different from zero, as it occurs in the purely quantum microcanonical ensemble.

Lastly, the proposed formalism can be easily extended to a continuous spectrum of energy. It is enough to consider in Eq. \eqref{eq:propRHmC}, instead of a sum, an integral over the relevant subset of $M_Q$ for the desired energy interval, as it is usually done in the purely quantum case \cite{Cohen-Tannoudji2020}.

\section{The microcanonical ensemble in the limit of small range of energies}
\label{sec:lim0}

In classical statistical mechanics, it is the case that the microcanonical ensemble, given for an energy interval $[E, E + \epsilon]$ by Eq. \eqref{eq:class_mC}, can be represented in the $\epsilon \to 0$ limit by a Dirac $\delta$ distribution $\delta \left( H_C(\xi)-E \right)$, where $H_C \in \mathcal{O}_C$ is the Hamiltonian function of the classical system. This limit is usually understood as a sequence where $\epsilon$ can be taken arbitrarily small, and the limit $\epsilon\rightarrow 0$ is well defined in the space $\mathcal{P}_C$ of probability distributions. The physical interpretation of this limit is that, if the resolution on the measurement of the energy of an isolated system increases arbitrarily, the microcanonical ensemble is always qualitatively the same, just with narrower energy localization.

In quantum mechanics this is not always the case. In particular, given the quantum microcanonical ensemble \eqref{eq:quant_mC} for an energy interval $[E, E + \epsilon]$ of a quantum system with Hamiltonian $H_Q \in \mathcal{O}_Q$, if $\epsilon < e_i - E$ for every $e_i \in {\rm spec}(H_Q)$, then the quantum microcanonical ensemble is empty, assigning null probability to all events. In fact, if the energy spectrum is discrete, for small enough $\epsilon > 0$ the quantum microcanonical ensemble is empty for almost all $E\in\mathbb{R}$; if $\epsilon$ is large enough to accommodate a single energy state, then the microcanonical ensemble only assigns nonzero probability to its eigenstates, while for large values of $\epsilon$ the microcanonical ensemble is an equiprobable mixture of eigenstates of the Hamiltonian whose eigenvalues lay within the energy range. This shows that the quantum microcanonical ensemble cannot be understood in general as a convergent sequence in the space of density states with similar qualitative features to that of classical statistics. Notice, though, that this analysis makes sense, only, for isolated (and hence idealized) systems. In such a case, assuming that it is possible to perform experiments with arbitrary precision and measuring suitable magnitudes as the variance of the energy, we should expect a discontinuous behavior when the value of $\epsilon$ selects several eigenspaces, just one, or none at all. Had we performed a similar sequence of experiments on a classical system, the variance of the energy would have changed continuously.

Hybrid statistics, while sharing many relevant characteristics with quantum statistics, nonetheless shows the same qualitative behavior as classical statistics in the limit of small energy values of the microcanonical ensemble in most relevant examples. To show this property, let us consider a simple hybrid system in which its classical variables $\xi = (R,P)$ represent classical positions $R$ and momenta $P$. Consider also that the hybrid Hamiltonian $\hat{H} \in \hat{O}_H$ contains purely classical terms incorporating a kinetic term $\frac{1}{2} P^2$ and a potential term $V(R)$, while the description of the quantum subsystem and its interaction with the classical part is contained in an additional term $H_Q(R)$ depending parametrically only on $R$:
\begin{equation}
\label{eq:HEjEps}
\hat H(R,P) = \left(\frac{1}{2} P^2 + V(R) \right) \mathbb{I} + H_Q(R), \quad (R,P) \in M_C,
\end{equation}
with $\mathbb{I}$ the identity operator.
Notice that the kinetic energy is unbounded from above this is a very common property of hybrid systems in applications and does not affect the states in the adiabatic basis.
For every $(R,P) \in M_C$, consider the spectrum ${\rm spec} (\hat H(R,P))$ of the hybrid Hamiltonian, its ground state $e_0(R,P)$ and its absolute minimum value $E_0 = \min_{(R,P) \in M_C}e_0(R,P)$. Then, for any $E \geq E_0$ and any $\epsilon >0$, consider the hybrid microcanonical ensemble in an interval $[E, E + \epsilon]$ defined by Eq. \eqref{eq:HmC}, which can also be written as
\begin{equation}
\label{eq:mC2}
\hat\rho_{\mu}(\xi) = \frac{1}{Z_\epsilon} \sum_{i=0}^{n^\xi} \sum_{j=1}^{N_i^\xi} \epsilon^{-1} \mathbf{1}_{[E,E+\epsilon]}(e_i^\xi) \pi_{i,j}^\xi,
\end{equation}
with $n^\xi$ the number of eigenvalues of $\hat{H}(\xi)$ at $\xi$ (which may be infinite), $Z_\epsilon = \epsilon^{-1} \Omega$ the normalization constant, depending on $\epsilon$, and $\mathbf{1}_A (r)$ the indicator function defined as
\begin{equation}
\mathbf{1}_A (r) = \begin{cases} 1 \mbox{ if } r \in A \\ 0 \mbox{ if } r \notin A \end{cases}, \quad A \subset \mathbb{R}, \ r \in \mathbb{R}.
\end{equation}
It is clear that the set of eigenstates of the adiabatic basis with energy within this interval is always nonempty regardless of $\epsilon$, as the kinetic term does not modify the basis but increases its energy by any arbitrary amount. If the $\epsilon \to 0$ limit is taken in Eq. \eqref{eq:mC2}, then
\begin{equation}
\label{eq:mC0}
\begin{gathered}
\hat\rho_{\mu}^0 (\xi) =
\lim_{\epsilon \to 0} \hat\rho_{\mu}(\xi)
= \frac{1}{Z_0} \sum_{i=0}^{n^\xi} \sum_{j=1}^{N_i^\xi} \delta \left( e_i^\xi-E \right) \pi_{i,j}^\xi, \\
Z_0 = \int_{M_C} \sum_{i=0}^{n^\xi} \delta \left( e_i^\xi-E \right) N_i^\xi d\mu_C,
\end{gathered}
\end{equation}
with the usual definition of Dirac $\delta$ distribution as the limit of an indicator function \cite{Stein2009}.

While the recovery of the $\delta$ function limit in the hybrid case is an interesting theoretical result, one may argue that this property is just a formal idealization  introduced in textbooks in the context of classical mechanics. However, we consider that this result has realistic physical implication not only in the value of the limit itself (which relates to the variance of the energy of the ensemble, and is a desirable property for a thermodynamical ensemble), but in the sequencial nature of the limit. The fact that the accuracy of measurements can be arbitrarily improved without qualitative changes is a requirement for the statistical theory to be consistent in its range of application. After all, being able to have compatible results in this sense under increasingly finer measurements is at the core of the appropriate definition of a microcanonical ensemble. In other words, we here prove that, even if different ensembles could be compatible with varying accuracy in the measurements, only the microcanonical ensemble is compatible with any possible accuracy.

In summary, the hybrid microcanonical ensemble for this system has the desirable property of having non trivial statistical behavior for all $\epsilon > 0$ and having the limit $\epsilon \to 0$ defined in a distributional sense.
Notice that, as degeneracy is taken into account, there exists for each $\xi \in M_C$ at most a single state with energy $E$. Thus, for every $\xi \in M_C$, either $\hat\rho_{\mu}^0 (\xi) = 0$ or it is proportional to a single Dirac $\delta$ distribution. However, as long as only a discrete and finite subset of $M_C$ describe hybrid states with energy $E$, the partition function $Z_0$ is simply
\begin{equation}
\label{eq:Z0}
Z_0 = \sum_{(i,\xi) \in \sigma_E} N_i^\xi, \quad \sigma_E = \left\{(i,\xi) \in \mathbb{N} \times M_C \mid e_i^\xi = E \right\}.
\end{equation}
In other words, $Z_0$ is the number of hybrid MESs with energy exactly equal to $E$.

More generally, instead of Eq. \eqref{eq:HEjEps}, consider a hybrid Hamiltonian whose spectrum varies continuously with the classical variables. For $\epsilon >0$, consider the hybrid microcanonical ensemble \eqref{eq:HmC} for an energy interval $[E, E + \epsilon]$. If there exists at least one state of the adiabatic basis with energy in this interval, and as long as the corresponding energy does not present a relative extreme within the interval, then  due to the continuity of the spectrum the parameter $\epsilon$ can be taken arbitrarily small, and in the $\epsilon \to 0$ limit the hybrid canonical ensemble will be different from zero. Again, assuming that experiments can be made with arbitrary precision, the variance of the total hybrid energy should exhibit a continious behavior when $\epsilon$ is reduced (for $\epsilon$ small enough so that only one eigenstate contributes at each $\xi$ at most).

We have thus shown that, in general cases, the hybrid microcanonical ensemble is non trivial in the $\epsilon\rightarrow 0$ limit. This property, present in classical systems but absent in discrete-spectrum quantum systems, is recovered in hybrid systems even if, for fixed classical variables, the spectrum of energies is discrete (i.e. the quantum subsystem is discrete), making the ensemble richer and allowing to interpret it as a sequence of arbitrarily small $\epsilon$, avoiding the trivial behavior of the quantum microcanonical ensemble in this cases.

\subsection{Limit classical marginal distribution}

Let us now consider the classical marginal distribution in the $\epsilon \to 0$ limit. In this shape, it is most evident that the microcanonical ensemble is the linear combination of tensor products of several classical (in terms of Dirac's $\delta$ distributions) and quantum (with $\epsilon$ small enough to include just one eigenstate in the case of discrete spectrum)  microcanonical ensembles. Let us show this by computing in the first the marginal classical distribution of \eqref{eq:mC0}
\begin{equation}
F_{\mu,C}^0(\xi) = \mathrm{Tr}\left(\hat\rho_{\mu}^0(\xi)\right) =
\frac{1}{Z_0} \sum_{i=0}^{n^\xi} \delta \left( e_i^\xi-E \right) N_i^\xi,
\end{equation}
where the degeneracy factor $N_i^\xi$ indicates the density of internal states due to the tracing-out of the quantum subsystem. As in Eq. \eqref{eq:mC0}, only one Dirac $\delta$ distribution can be different from zero at each $\xi \in M_C$. Notice that, for trivial quantum systems of dimension 1, degeneracy is $N_i^\xi=1$ for all states $i$ and all $\xi \in M_C$. In this situation, the classical marginal distribution $F_{\mu,C}^0(\xi)$ reproduces the classical microcanonical ensemble, showing a good classical limit behavior.

\subsection{Quantum distributions in the limit}

However, the quantum conditional ensemble $\tilde\rho_{\mu,Q,\xi}^0$ in the $\epsilon \to 0$ limit can be simply identified from the relation \eqref{eq:defRhoH} applied to the $\epsilon \to 0$ limit:
\begin{equation}
\begin{aligned}
\hat\rho_{\mu}^0 (\xi) = & F_{\mu,C}^0(\xi) \tilde\rho_{\mu,Q,\xi}^0,  \quad \xi \in M_C \\
\Rightarrow & \tilde\rho_{\mu,Q,\xi}^0 = \begin{cases}
\displaystyle\frac{1}{N_i^\xi} \sum_{j=1}^{N_i^\xi} \pi_{i,j}^\xi, \quad \exists ! i \in \mathbb{N} \mid e_i^\xi = E, \\
0, \quad \nexists i \in \mathbb{N} \mid e_i^\xi = E,
\end{cases}
\end{aligned}
\end{equation}
where the fact that, at every $\xi \in M_C$, there exists at most one state in the set of MESs with energy exactly equal to $E$ is taken into account. The quantum conditional ensemble is similar to the quantum microcanonical ensemble with energy $E$. However, unlike the purely quantum case, the quantum conditional ensemble for a generic hybrid system may be well defined (i.e. different from zero for some $\xi \in M_C$) for a continuous range of energies. In particular, if the Hamiltonian incorporates a kinetic term such as in Eq. \eqref{eq:HEjEps}, then for every energy above the minimum the quantum conditional ensemble is non zero for at least one $\xi \in M_C$. This new property is shared by the quantum marginal distribution in the $\epsilon \to 0$ limit
\begin{equation}
\rho_{\mu,Q}^0 =
\int_{M_C} \hat\rho_{\mu}^0 (\xi) d \mu_C = \frac{1}{Z_0} \sum_{(i,\xi) \in \sigma_E} \sum_{j=1}^{N_i^\xi} \pi_{i,j}^\xi,
\end{equation}
with $Z_0$ and $\sigma_E$ defined in Eq. \eqref{eq:Z0}. For realistic hybrid systems, i.e those with unbounded kinetic or potential terms in the Hamiltonian as in Eq. \eqref{eq:HEjEps}, the quantum marginal distribution $\hat\rho_{\mu,Q}^0$ is a convex combination of $Z_0$ rank-one projectors $\pi_{i,j}^\xi$ for every energy above the minimum. In the trivial case that eigenvalues and eigenstates do not vary with $\xi$, the quantum marginal distribution becomes the microcanonical ensemble of a purely quantum system with energy $E$, which as usual is only defined if $E$ is an eigenvalue of the Hamiltonian.

\section{Relation between hybrid microcanonical and canonical ensembles}
\label{sec:HybridCan}

Microcanonical and canonical ensembles of hybrid quantum-classical systems are directly related. The canonical ensemble of a statistical system can obtained as the marginal distribution of a system interchanging energy with a reservoir when both are considered together as a single isolated system in equilibrium, hence described by the microcanonical ensemble.

\subsection{MESs for the  compound hybrid system}

Consider a hybrid system $S$, with hybrid phase space $M^S = M_C^S \times M_Q^S$ and $\mathcal{H}^S$ as the Hilbert space defining the quantum subsystem phase space $M_Q^S$, and a reservoir $R$, which we assume purely quantum for simplicity, hence with phase space $M^R = M_Q^R$ and Hilbert space $\mathcal{H}^R$. The compound system S+R is an isolated hybrid system with phase space $M = M_C \times M_Q$, with $M_C = M_C^S$ and where the quantum subspace $M_Q$ corresponds to the rank-1 projectors onto the tensor product Hilbert space $\mathcal{H} = \mathcal{H}^S \otimes \mathcal{H}^R$. Being isolated, and without further restrictions, the state of the compound system S+R is described by a hybrid microcanonical ensemble \eqref{eq:HmC} within a certain energy interval $[E,E+\epsilon]$.

Consider the respective Hamiltonians of the two subsystems: a hybrid observable $\hat{H}^S \in \hat{\mathcal{O}}_H$ for the system $S$, and a purely quantum observable $H^R \in \mathcal{O}_Q$ for the reservoir $R$, with the following eigenvalue equations:

\begin{equation}
\hat{H}^S(\xi) |u_{i,j}^{S\xi} \rangle = e_{i}^{S\xi} |u_{i,j}^{S\xi} \rangle, \ H^R |v_{m,n}^R \rangle = e_{m}^R |v_{m,n}^R \rangle.
\end{equation}
The eigenvectors, assumed to be normalized, conform an adiabatic bases $b^S(\xi) = \{|u_{i,j}^{S\xi} \rangle \}$ for the hybrid system $S$ and a basis $b^R = \{ |v_{m,n}^R \rangle \}$ for $\mathcal{H}^R$.
If a weak coupling between the system $S$ and the reservoir $R$ is assumed, then each state described by 
$|u_{i,j}^{S\xi},v_{m,n}^R \rangle= |u_{i,j}^{S\xi} \rangle \otimes |v_{m,n}^R \rangle$
has energy $e_{i}^{S\xi} + e_{m}^R$. 
The set $b(\xi) = \{|u_{i,j}^{S\xi} \rangle \otimes |v_{m,n}^R \rangle \rangle \}$ is then an adiabatic basis for the whole hybrid system $S+R$.
Thus, a maximal set of MESs for the compound hybrid system with known energy, in the fashion of Eq. \eqref{eq:MEEs_H}, can be written as follows:
\begin{equation}
\begin{aligned}
\mathcal{B}_{S+R} = & \left\{ \left(\xi, \pi^\xi_k \right) \in M \mid k = (i,j,m,n) \in \mathbb{R}^4, \right. \\
& \qquad \left. \pi^\xi_k =  |u_{i,j}^{S\xi},v_{m,n}^R \rangle \langle u_{i,j}^{S\xi},v_{m,n}^R | \right\}.
\end{aligned}
\end{equation}
The ``weak coupling'' condition is normally justified in terms of the size of system and reservoir: first of all, the reservoir is very large in comparison with the system (a condition that becomes important in the following step); furthermore, both system and reservoir are large enough that the frontier region between them becomes unimportant, which is what permits to neglect the interaction term.

\subsection{Microcanonical ensemble of the compound system}

The microcanonical ensemble of the compound system is described by restricting the energy of states in $\mathcal{B}_{S+R}$ to the energy interval $[E,E+\epsilon]$, obtaining as in Eq. \eqref{eq:subMESs} the following subset:
\begin{equation}
\begin{aligned}
\mathcal{A}_{S+R}^{E, \epsilon} = & \left\{ (\xi, \pi^\xi_k ) \in \mathcal{B}_{S+R} \mid k = (i,j,m,n) , \right. \\
& \qquad \left. e_{i,j}^{S\xi} + e_{m,n}^R \in [E, E + \epsilon] \right\}.
\end{aligned}
\end{equation}
As long as $\mathcal{A}_{S+R}^{E, \epsilon} \neq \emptyset$, the microcanonical ensemble of the compound system is given by Eq. \eqref{eq:HmC}:
\begin{equation}
\hat{\rho}_\mu^{S+R} (\xi) = \frac{1}{\Omega} \left\{ \begin{aligned}
& \sum_{k \in \iota(\xi)} \pi_k^\xi, && \iota(\xi) \neq \emptyset \\
& 0, && \iota(\xi) = \emptyset
\end{aligned} \right., \quad \xi \in M_C^S,
\end{equation}
where the normalization factor $\Omega$, computed in Eq. \eqref{eq:Omega}, represents the number of states of the compound system $S+R$ whose energy is within the interval $[E, E + \epsilon]$, i.e. the number of states in $\mathcal{A}_{S+R}^{E, \epsilon}$.

\subsection{Connection between the canonical and microcanonical ensembles}

Once that the full compound system $S+R$ is described, our interest is to analyze the resulting ensemble for the hybrid system $S$. To do so, we will work under the assumption that the reservoir is large enough so that, for any $\xi \in M_C^S$ such that $\iota(\xi) \neq \emptyset$, there exists at least one state of the compound system with energy $e^T = e^S + e^R \in [E, E + \epsilon]$. The marginal hybrid density matrix can thus be obtained by tracing out the quantum degrees of freedom corresponding to the reservoir $R$ \cite{Cohen-Tannoudji2020}, which can be computed directly as the projectors of the compound system $S+R$ are separable:
\begin{widetext}
\begin{equation}
\label{eq:TrR}
\begin{aligned}
& \hat{\rho}^S (\xi)
= \mathrm{Tr}_R \left( \hat{\rho}_\mu^{S+R} (\xi) \right)
= \sum_{|v_{m,n}^R\rangle \in b^R} \left( \mathbb{I}^S \otimes \langle v_{m,n}^R | \right) \hat{\rho}_\mu^{S+R} (\xi) \left( \mathbb{I}^S \otimes | v_{m,n}^R \rangle \right) \\
& = \frac{1}{\Omega} \sum_{|v_{m,n}^R\rangle \in b^R} \sum_{k \in \iota(\xi)} \left( \mathbb{I}^S \otimes \langle v_{m,n}^R | \right)
\left|u_{i,j}^{S\xi},v_{m,n}^R \right\rangle \left\langle u_{i,j}^{S\xi},v_{m,n}^R \right| \left( \mathbb{I}^S \otimes | v_{m,n}^R \rangle \right) = \frac{1}{\Omega} \sum_{k \in \iota(\xi)} \Omega^R_i \left|u_{i,j}^{S\xi} \right\rangle \left\langle u_{i,j}^{S\xi} \right|, \quad \iota(\xi) \neq \emptyset,
\end{aligned}
\end{equation}
\end{widetext}
with $\mathbb{I}^S$ the identity operator on $\mathcal{H}^S$ and $\Omega^R_i$ the number of states of the reservoir whose energy is $e^T-e^{S,\xi}_{i}$, for each possible $i$ at $\xi \in M_C$ and every $e^T \in [E, E + \epsilon]$. 

Assuming $\epsilon$ small enough, each $\Omega^R_i$ is approximately equal to the number of states of the reservoir $R$ with energy exactly equal to $E-e^{S,\xi}_{i}$, as shown in Eq. \eqref{eq:Omega}. The number of states of a system with a specific value of energy can be computed as the exponential of the entropy of a microcanonical ensemble at said energy (taking the Boltzmann constant equal to 1). Thus, using the von Neumann's entropy for a microcanonical ensemble $\hat{\rho}_\mu^{R}$ of the quantum reservoir $R$ at energy $e_R$, written as $S_N^R(e_R) = S_N[\hat{\rho}_\mu^{R}]$, the number of states with said energy is $\exp \left( S^R_N (e_R) \right).$ If we further assume that only a small part of the total energy corresponds to the system $S$, i.e. $e^{S,\xi}_{i} \ll E$ for every $i$ and every $\xi \in M_C$, then we can consider the following power expansion with respect to each $e^{S,\xi}_{i}$ truncating to first order:
\begin{equation}
\label{eq:SSeries}
S^R_N (E-e^{S,\xi}_{i}) \simeq S^R_N (E) - \beta e^{S,\xi}_{i}, \quad
\beta = \frac{dS_N^R}{de_R} (E).
\end{equation}
Notice that $\beta$ is a constant, the value of the derivative of the entropy of the reservoir $S_N^R$ with respect to its energy at $e_R = E$. Taking the exponential of the entropy, the value of $\Omega^R_i$ is approximated as
\begin{equation}
\label{eq:OmegaSeries}
\Omega^R_i \simeq \exp \left( S_N^R (E-e^{S,\xi}_{i}) \right) \simeq \exp \left( S_N^R (E) \right) \exp \left( - \beta e^{S,\xi}_{i} \right).
\end{equation}
With this approximation, the whole volume $\Omega$ can be written as
\begin{align}
\Omega & = \int_{M_C} d \mu_C \sum_i \Omega^R_i N^\xi_i \nonumber \\
& \simeq \exp \left( S_N^R (E) \right) \int_{M_C} d \mu_C \sum_i \exp \left( - \beta e^{S,\xi}_{i} \right) N^\xi_i \nonumber\\
& = \exp \left( S_N^R (E) \right) \int_{M_C} d \mu_C \, \mathrm{Tr} \left( \exp \left(-\beta \hat{H}^S(\xi)\right)\right),
\end{align}
with $N^\xi_i$ the degeneracy at $\xi \in M_C$ of the $i$th eigenvalue of the Hamiltonian $\hat{H}^S(\xi)$.

Substituting in Eq. \eqref{eq:TrR}, the marginal ensemble for system $S$ can be obtained. Additionally, under the $e^{S,\xi}_{i} \ll E$ assumption for every $i$, every state of system $S$ can be populated, so that the sum in Eq. \eqref{eq:TrR} extends over all possible states $i,j$. In conclusion, the marginal ensemble of system $S$ is given, up to first order, by the following hybrid density matrix:
\begin{equation}
\label{eq:can}
\begin{aligned}
\hat{\rho}^S (\xi)
& \simeq \frac{1}{Z} \sum_{i,j} \exp \left( - \beta e^{S,\xi}_{i} \right) \left|u_{i,j}^{S\xi} \right\rangle \left\langle u_{i,j}^{S\xi} \right| \\
& = \frac{\exp \left(-\beta \hat{H}^S(\xi)\right)}{Z}, \\
Z & = \int_{M_C} d \mu_C \, \mathrm{Tr} \left( \exp \left(-\beta \hat{H}^S(\xi)\right)\right).
\end{aligned}
\end{equation}

This is, as expected, the hybrid canonical ensemble for a hybrid system in contact with a reservoir, as first introduced in Ref. \cite{Alonso2020}. Within this interpretation, the constant $\beta$, introduced in Eq. \eqref{eq:SSeries} without any physical interpretation at the moment, can now be identified as the inverse of the temperature $T = \beta^{-1}$ of the reservoir. The partition function $Z$ is derived from the approximation \eqref{eq:SSeries}, and allows for the normalization condition \eqref{eq:normRH} to be satisfied.

The results show the connection between the hybrid microcanonical and canonical ensembles, a required property of the proposed ensemble that supports is validity as its extension to hybrid systems of the standard purely quantum and classical ensembles. It is immediate to show with similar arguments that, if considering either classical or hybrid reservoirs instead of quantum, then one also obtains the hybrid canonical ensemble as the marginal distribution of system $S$, the only difference being the computation of the partial trace, which in every case provides as a factor the number of states with energy $e^T-e^{S,\xi}_{i}$.

\section{Microcanonical ensemble of a hybrid qubit}
\label{sexExample}

Let us consider a simple interaction model  consisting of a two-level quantum system ($\mathcal{H} = \mathbb{C}^2$) coupled to a classical bath with a single parameter $R \in \mathbb{R} = M_C$, with hybrid Hamiltonian given by the following expression:
\begin{equation}
\hat{H}(R) = V(R) \mathbb{I} + H_Q + \hat{H}_{\rm int} (R),
\end{equation}
where $V(R)$ represents the potential of the classical bath, $\mathbb{I}$ is the identity operator on $\mathcal{H}$, $H_Q$ is the Hamiltonian of the quantum subsystem and $\hat{H}_{\rm int} (R)$ describes the classical-quantum interaction. The model has been used  \cite{Nielsen2000a} (see also Ref. \cite{Legget1987}) as an effective model for a dissipative quantum system where only two energy levels have physical relevance (although it can be generalized to higher finite dimensional cases). Notice that, from a hybrid system point of view, it is a particularly simple case, since classical degrees of freedom are not dynamical for they are acting as a bath for the quantum ones. Nonetheless, our MaxEnt construction is not dependent on the system dynamics, and therefore we decided to choose this simple yet interesting model to exemplify our construction without computational difficulties.

To obtain simple but nontrivial results, we will use a quadratic bath potential and a simple quantum Hamiltonian $H_Q$ with eigenvalues $0$ and $0.1$ in natural units ($\hbar = 1$); their expressions in a basis of eigenvectors of $H_Q$ are
\begin{equation}
V(R) = R^2, \
H_Q = \begin{pmatrix}
0 & 0 \\ 0 & 0.1
\end{pmatrix}, \ \hat{H}_{\rm int} (R) = R \begin{pmatrix}
0 & 1 \\ 1 & 0
\end{pmatrix}.
\end{equation}

The solutions to the eigenvalue equation for the hybrid Hamiltonian $\hat{H}(R)$ can be computed analytically for every $R \in \mathbb{R}$: its expression for $R \neq 0$ are
\begin{equation}
\label{eq:ejEigR}
\begin{gathered}
\hat{H}(R) |u^R_\pm\rangle = e^R_\pm |u^R_\pm\rangle; \quad
e^R_\pm = R^2 + \frac{1}{20} \pm \frac{\sqrt{400 R^2 + 1}}{20}; \\
|u^R_\pm \rangle = \frac{1}{\sqrt{v^R_\pm}}  \begin{pmatrix} \sqrt{400 R^2 + 1} \mp 1 \\ \pm 20 R \end{pmatrix}; \\
v^R_\pm = 800 R^2 + 2 \mp 2 \sqrt{400 R^2 +1};
\end{gathered}
\end{equation}
while, for $R=0$,
\begin{equation}
\label{eq:ejEig0}
\begin{gathered}
\hat{H}(0) |u^0_\pm\rangle = e^0_\pm |u^0_\pm\rangle; \quad
e^0_- = 0, \ e^0_+ = \frac{1}{10}; \\
|u^0_- \rangle = \begin{pmatrix} 1 \\ 0 \end{pmatrix}, \ |u^0_+ \rangle = \begin{pmatrix} 0 \\ 1 \end{pmatrix}
\end{gathered}.
\end{equation}

\begin{figure*}[t]
\centering
\includegraphics[width=\textwidth]{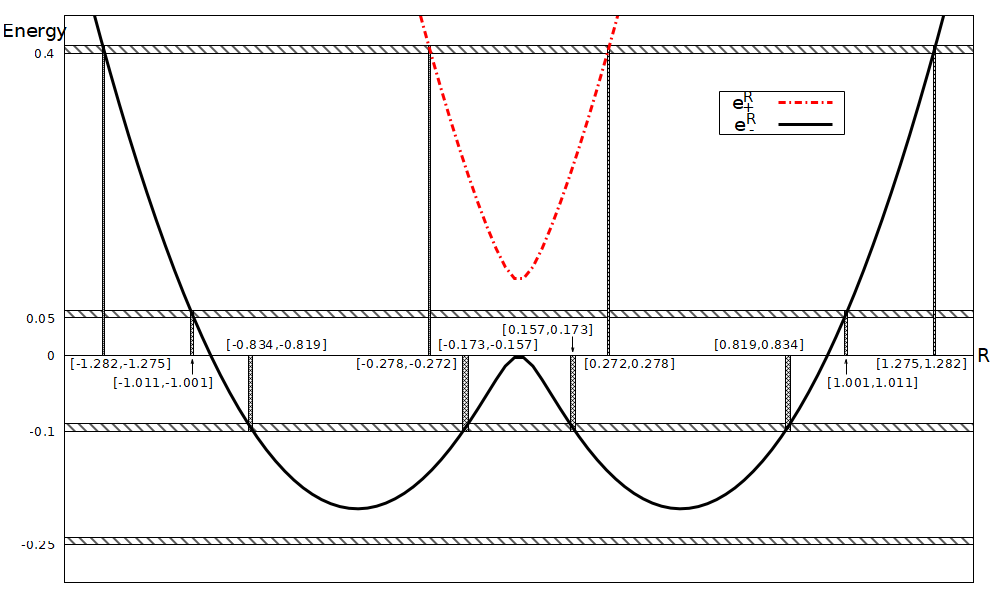}
\caption{Eigenvalues $e_+^R$ and $e_-^R$ of the hybrid Hamiltonian $\hat{H}(R)$, shown in Eqs. \eqref{eq:ejEigR} and \eqref{eq:ejEig0}, are represented as functions of the classical parameter $R$. Additionally, energy bands of width $\epsilon = 0.01$ are represented horizontally, with bottom energies equal to $-0.25$, $-0.1$, $0.05$, $0.4$. Each intersection of the energy bands with the curves of the eigenvalues is projected vertically onto the $R$ axis, defining an interval of values, as shown along the axis. These intervals have been computed numerically from Eq. \eqref{eq:ejEigR}.}
\label{fig:eig}
\end{figure*}

Both eigenvalues are represented with respect to the classical parameter $R$ in Fig.~\ref{fig:eig}. These quantities represent the energy of the whole hybrid system when its state is $(R, |u^R_- \rangle \langle u^R_-|)$ and $(R, |u^R_+ \rangle \langle u^R_+|)$, respectively. Therefore, the eigenvalues give us the necessary information to define the hybrid microcanonical ensemble at different energies.

Figure~\ref{fig:eig} shows what happens when we consider a fixed value $E$ for the energy of the hybrid system. Each horizontal band represents a different interval of energy values of the form $[E, E+ \epsilon$], with $\epsilon = 0.01$ in all cases. By analyzing the intersections between these bands and the eigenvalues, we can determine which states of the whole hybrid system have an energy contained in the corresponding interval. Depending on the value of the energy, it is possible that none, one or both of the energy levels of the quantum subsystem are involved. Let us discuss the different cases shown in the figure.
\begin{itemize}
\item For the lower band, corresponding to $E=-0.25$, there is no intersection with the curves $e^R_\pm$. Indeed, the hybrid system has a minimum energy of approximately $-0.2$, as shown in the figure. Therefore, it is not possible to build a microcanonical ensemble with energy below this value.
\item For $E=-0.1$, the corresponding band intersects four times the curve $e^R_-$. Each intersection occurs for an interval of $R$ values, as indicated in the figure. For any $R$ in those intervals, the energy of the hybrid state lays in the interval $[-0.1, -0.09]$, therefore contributing to the microcanonical ensemble. As discussed above, all this states are mutually exclusive, as they have different values of $R$. No state involving $e^R_+ $ contributes to the ensemble, as their energy is always above $-0.1$. Thus, the microcanonical ensemble for energies in the interval $[-0.1, -0.09]$ is
\begin{equation}
\hat{\rho}_{\mu} (R) = \left\{ \begin{array}{l}
\displaystyle \frac{1}{0.062} |u^R_-\rangle \langle u^R_-|,\\[10pt]
\quad \mbox{if } |R| \in [0.157, 0.173] \cup [0.819, 0.834], \\[5pt]
0, \ \mbox{otherwise},
\end{array} \right.
\end{equation}
where the normalization factor $0.062$ is the total length of the interval of $R$ values such that ${\hat{\rho}_{\mu} (R) \neq 0}$.
\item For $E=0.05$, the corresponding band again only intersects the curve $e^R_-$, but only twice this time, due to its curvature. The microcanonical ensemble for energies in the interval $[0.05, 0.06]$ is
\begin{equation}
\label{eq:ex2}
\hat{\rho}_{\mu} (R) = \left\{ \begin{array}{ll}
\displaystyle \frac{1}{0.02} |u^R_-\rangle \langle u^R_-|, & \mbox{if } |R| \in [1.001,1.011],\\ \\
0 & \mbox{otherwise}.
\end{array} \right.
\end{equation}
\item Finally, let us consider the case $E=0.4$. Figure~\ref{fig:eig} shows that the band of energies $[0.4, 0.41]$ intersects both curves $e^R_-$ and $e^R_+$. Therefore, there are contributions of both quantum states to the microcanonical ensemble; however, each one occurs for different values of $R$, as seen in the figure. In consequence, the hybrid microcanonical ensemble for energies in the interval $[0.4, 0.41]$ is
\begin{equation}
\hat{\rho}_{\mu} (R) = \left\{ \begin{array}{ll}
\displaystyle \frac{1}{0.026} |u^R_-\rangle \langle u^R_-|, & \mbox{if } |R| \in [1.275,1.282],\\ \\
\displaystyle \frac{1}{0.026} |u^R_+\rangle \langle u^R_+|, & \mbox{if } |R| \in [0.272,0.278],\\ \\
0 & \mbox{otherwise}.
\end{array} \right.
\end{equation}
Notice that the normalization factor $0.026$ is included in both nonzero pieces of the ensemble and has to be computed considering all the values of $R$, as it is necessary for the normalization of $\hat{\rho}_{\mu}(R)$ by Eq. \eqref{eq:normRH}.
\end{itemize}

As a last remark, notice that, in connection to what was exposed in Sec. \ref{sec:lim0}, the $\epsilon \to 0$ limit of all microcanonical ensembles described in the example are well defined and share the same properties as a classical microcanonical ensemble. For example, consider the case $E = 0.05$. If the $\epsilon \to 0$ limit is taken in ensemble \eqref{eq:ex2}, then one simply gets
\begin{equation}
\begin{aligned}
& \hat{\rho}_\mu (R) = \delta \left( R- R_+ \right) |u_-^{R_+} \rangle \langle u_-^{R_+} | \\
& + \delta \left( R- R_- \right) |u_-^{R_-} \rangle \langle u_-^{R_-} |, \quad R_+ = - R_- = 1.001,
\end{aligned}
\end{equation}
as the lower eigenvalue $e_-^R$ takes value $0.05$ at ${R = \pm 1.001}$. Similar limits can be computed in all other cases.

\section{Conclusions and future work}
\label{sec:conclusions}

In this work, we present a detailed description of the phase space of hybrid quantum systems and the description of statistics. The mathematical framework allows use to completely characterize probability ensemble by means of hybrid density matrices. Within this context, we also perform a computation of hybrid quantum-classical microcanonical ensembles based on a maximum entropy principle applied to the hybrid entropy. The resulting ensemble presents the expected characteristic of microcanonical ensemble, being an equiprobable distribution of MESs with energies in the allowed interval. Furthermore, the classical and quantum microcanonical ensembles are obtained in the corresponding limit cases.

One of the most remarkable properties of the hybrid microcanonical ensemble here presented is its ability to propagate classical properties to its quantum components. Typically, classical microcanonical ensembles can be defined for any fixed energy of the system above its minimum. The same is not true purely quantum systems, as a microcanonical ensemble can only be strictly formulated for energies equal to the eigenvalues of the Hamiltonian, whose spectrum is generally discrete; although, naturally, for large systems any feasible small range of energies always contains some eigenvalues and the discreteness of the spectrum is not a problem in practice. We show in this article how, given a hybrid microcanonical ensemble, its marginal quantum ensemble is similar to a purely quantum microcanonical ensemble but, due to the influence of the classical subsystem, can be generally defined for a continuum range of energies. This extension of classical properties to quantum statistics may be key to the description of a framework in which the microcanonical ensemble can be intrinsically applied to any system, without the necessity for additional properties such as large systems, and for this reason may be useful in the analysis of more general systems than those proposed in this article, such as molecules and quantum gravity models.

We also show how the hybrid microcanonical ensemble allows us to compute the hybrid canonical ensemble proposed in Ref. \cite{Alonso2020}. By considering the microcanonical ensemble of a compound system with a hybrid subsystem and a quantum reservoir, and using the typical assumptions in this setting \cite{Cohen-Tannoudji2020}, the hybrid canonical ensemble is obtained as the marginal ensemble of the hybrid subsystem. This is another proof of the validity of the proposed hybrid microcanonical ensemble in the context of building a self-coherent statistical description of hybrid quantum-classical systems.

Notice that our analysis did not consider the dynamics of the microstates at all. However, for the maximum entropy principle to produce physically meaningful thermodynamic ensembles, it is well known that a suitable microdynamics must be defined having the maximum entropy solution as an equilibrium  state. The description of appropriate hybrid quantum-classical dynamics is an open problem that we will analyze in future articles. Potential applications to the definition of suitable hybrid thermostats also require of a suitable dynamics, which, hopefully, are ergodic on the submanifold of fixed energy. In such a case time averages over single trajectories, which are easy to implement numerically, would provide us with statistical averages over this microcanonical ensemble. Using these averages with the microcanonical-canonical relation discussed above, fixed temperature averages (always difficult numerically) would be easily implemented.

Our work can be related with other existing approaches to hybrid systems. For example, in Ref. \cite{Hall2016} the author focus their attention on aspects of hybrid quantum-classical ensembles which do not require introducing the entropy. However, an approach based on configuration space ensembles overcomes many difficulties arising in other approaches, as explained in Sec. 8.1 of said reference.

Different dynamical approaches may also be considered by using new notions which generalize the idea of quantum operations (also known as quantum dynamical maps \cite{Sudarshan1961}) to the hybrid framework. Very recently, one possible generalization has been introduced (see Eq. (7) in Ref. \cite{Camalet2025}). Approaches based on a hybrid generalization of Koopman classical dynamics \cite{BouthelierMadre2023, Bondar2019,Gay-Balmaz2022, Gay-Balmaz2023}, or nonlinear approaches based on the restriction of Lioville Ehrenfest dynamics to a finite number of quantum moments \cite{Alonso2023b} may offer interesting candidates for hybrid equilibrium dynamics. Also, recent works deal with the description of out-of-equilibrium thermodynamics of hybrid quantum-classical systems \cite{Eglinton2024, Layton2025}, which provide a fertile ground for the application of entropy-based analysis of hybrid systems. Other relevant works deal with different aspects of dynamics of hybrid systems, such as path integrals \cite{Oppenheim2025, Layton2024}. We hope to expand our analysis and develop our formalism to incorporate all of these cutting-edge developments in a complete mathematical framework.

Future works will expand on this formulation and its applications. The development of a mathematically cohesive formalism of statistical hybrid systems is fundamental for the study of numerous systems, and the results of these mathematical studies may help to develop new and more efficient computational tools for the study of molecular and condensed matter models.


\begin{acknowledgments}
The authors acknowledge partial financial support of Grants No. PID2021-123251NB-I00 (JLA, AC, JC-G), No. CPP2021-008644 (CB), and No. PID2021-122961NB-I00 (JAJ-G) funded by Ministerio de Ciencia e Innovación (MCIN/AEI/10.13039/501100011033) and by the European Union (Next Generation EU), and of Grants No. E48\_23R (AC, JC-G) and No. E24\_23R (JAJ-G) funded by Gobierno de Aragón. Carlos Bouthelier-Madre acknowledges partial financial support to the project CPP2021-008644, funded by the Spanish Ministry of Science and Innovation through the Public-Private Collaboration 2021 call (State Plan for Scientific, Technical and Innovation Research 2021–2023), within PRTR and co-funded by the European Union - NextGenerationEU. We thank Profs. Marcel Reginatto, Sébastien Camalet, and Adrián Budini for their illuminating remarks in our discussions.

All authors contributed equally to this work.
\end{acknowledgments}

\section*{Data availability}
No data were created or analyzed in this study.


\end{document}